# CONSISTENCY PROBLEMS FOR JUMP-DIFFUSION MODELS

ERHAN BAYRAKTAR, LI CHEN, AND H. VINCENT POOR

ABSTRACT. In this paper consistency problems for multi-factor jump-diffusion models, where the jump parts follow multivariate point processes are examined. First the gap between jump-diffusion models and generalized Heath-Jarrow-Morton (HJM) models is bridged. By applying the drift condition for a generalized arbitrage-free HJM model, the consistency condition for jump-diffusion models is derived. Then we consider a case in which the forward rate curve has a separable structure, and obtain a specific version of the general consistency condition. In particular, a necessary and sufficient condition for a jump-diffusion model to be affine is provided. Finally the Nelson-Siegel type of forward curve structures is discussed. It is demonstrated that under regularity condition, there exists no jump-diffusion model consistent with the Nelson-Siegel curves.

We are indebted to Erhan Çinlar and Damir Filipović for helpful discussions.

1. The Arbitrage-free Condition for Generalized HJM Models

The purpose of this paper is to study consistency problems for multi-factor jump-diffusion term structure models of interest rates. The concept of consistency in this context was first introduced and studied in [4]. Previous works ([13], [14], [15]) in this area have focused on diffusion models without considering jumps. Because jump-diffusion models usually provide a better characterization of the randomness in financial markets than do diffusion models (see [1], [19]), there has been an upsurge in the modeling of interest rate dynamics with jumps (e.g. [3], [12], [17], [20]). Therefore it is of interest to clarify the consistency conditions for jump-diffusion models.

Consider a Heath-Jarrow-Morton (HJM) model ([16]) incorporating a marked point process. The dynamics of the forward curve for such a model can be given by

$$(1.1) \qquad dr(t,T) = \alpha(t,T)dt + \sigma(t,T)dB_t + \int_\Theta \rho(t,T,y)\mu(dt,dy),$$

where $B$ is a standard Brownian motion and $\mu(dt,dy)$ is a random measure on $\mathbb{R}_+ \times \Theta$ with the compensator $\nu(t,dy)dt$. Thus the price of a zero-coupon bond can be written as

$$(1.2) \qquad P(t,T) = e^{-\int_t^T r(t,u)du}.$$

A measure $\mathbb{Q}$ is said to be a local martingale measure if the discounted bond price

$$D(t,T) = \frac{P(t,T)}{e^{-\int_0^t r(s,s)ds}}$$

is a $\mathbb{Q}$-local martingale, for each $T \in \mathbb{R}_+$. It is well known that the existence of an equivalent local martingale measure implies the absence of arbitrage (e.g. see [9]).

Under regularity conditions, Björk et al. [5] give the following lemma for the arbitrage-free condition of a generalized HJM model defined by (1.1).

**Lemma 1.1.** *An equivalent local martingale measure exists if and only if the forward rate dynamics under this measure specified by (1.1) satisfy the following relation for $\forall\, 0 \leq t < T$.*

$$(1.3) \qquad \alpha(t,T) = \sigma(t,T) \int_t^T \sigma(t,s)ds - \int_\Theta \rho(t,T,y)e^{-\int_t^T \rho(t,u,y)du}\nu(t,dy).$$

*Proof.* See [5], Theorem 3.13 and Proposition 3.14. □

Lemma 1.1 gives the drift condition for a generalized HJM model, which generalizes the traditional arbitrage-free condition for diffusion HJM models. This result provides a way for us to derive consistency conditions for a multi-factor model with jumps.

In this paper, we consider the consistency issue for jump-diffusion models. In particular, in Section 2, we bring a jump-diffusion model into the generalized HJM framework, and derive the consistency condition for the coefficient functions of the model. In Section 3, we discuss a class of separable term structure models, which includes affine and quadratic models as special cases. In particular, the affine term structure is investigated and sufficient and necessary conditions for a jump-diffusion model to be affine are derived. A typical non-separable term structure model, namely the Nelson-Siegel term structure, is examined in Section 4. Under regularity conditions, we conclude there exists no jump-diffusion model consistent with the Nelson-Siegel forward curves. Brief concluding remarks are made in Section 5.

1.1. **Basic Notation.** Table 1 defines notation that will be used frequently in this paper.

2. CONSISTENCY CONDITIONS FOR MULTI-FACTOR JUMP-DIFFUSION MODELS

Consider a multi-factor term structure model $\mathcal{M}$ with forward rates of the form:

$$r(t,T) = G(T-t, X_t),$$

where the state process $X$ follows a jump-diffusion in which the jump part follows a multivariate point process with state space $(D, \mathcal{D})$ under a complete filtered probability space $(\Omega, \mathcal{F}, (\mathcal{F}_t), \mathbb{P})$:

$$(2.1) \qquad dX_t = b(X_t)dt + c(X_t)dW_t + \int_E N(dt, dz)k(X_{t^-}, z), \quad X_0 = x,$$

where $G : \mathbb{R}_+ \times D \mapsto \mathbb{R}$ is a deterministic function, $W_t$ is an $n$-dimensional standard Brownian motion and $N(\cdot, \cdot)$ is a Poisson random measure on $\mathbb{R}_+ \times E$ with mean measure $Leb \times \psi$. $\mathbb{P}$ denotes the equivalent martingale measure.

We will take $X$ in (2.1) to be piece-wise continuous. This implies that $\psi$ is a finite measure on $E$, i.e., $\psi(E) < \infty$ and that $\int_0^t ds \int_E \psi(dz) k(X_{s^-}, z) < \infty$ a.s., for each $t \in \mathbb{R}_+$. Moreover, later in this section and in the following sections we will restrict $X$ to be a jump diffusion process whose jumps have a fixed probability distribution $Q$ on $J$ and arrive with intensity $\{\lambda(X_t) : t \geq 0\}$ (see Assumption 2.2).

**Definition 2.1.** *A multi-factor jump-diffusion model $\mathcal{M}$ is said be consistent if the induced dynamics of the forward rates $G$ satisfies the relation (1.3), for all $x \in D$.*

**Assumption 2.1.** For the sake of analytical tractability, we make the following assumptions.

- *The function $G \in C^{1,2}(\mathbb{R}_+ \times D)$;*
- *the functions $b : D \mapsto \mathbb{R}^n$, $c : D \mapsto \mathbb{R}^{n \times n}$ and $k : D \times E \mapsto J \subset \mathbb{R}^n$ are deterministic and continuous; and*
- *(2.1) has a unique strong solution $X_t$ in $D$, for each $x \in D$.*

By Itô's formula, the dynamics of the forward rates $G(\cdot, X_t)$ can be derived as follows.

$$
\begin{aligned}
(2.2)\ G(\cdot, X_t) =\ & G(\cdot, x_0) + \int_0^t \partial_\tau G(\cdot, X_s) ds + \sum_{i=1}^n \int_0^t \partial_{x_i} G(\cdot, X_{s-}) b_i(X_{s-}) ds \\
& + \sum_{i=1}^n \int_0^t \partial_{x_i} G(\cdot, X_{s-}) \sum_{j=1}^n c_{i,j}(X_{s-}) dW_t^j \\
& + \sum_{i,j=1}^n \int_0^t a_{i,j}(X_{s-}) \partial_{x_i} \partial_{x_j} G(\cdot, X_{s-}) ds \\
& + \int_0^t \int_E N(ds, dz)[G(\cdot, X_{s-} + k(X_{s-}, z)) - G(\cdot, X_{s-})],
\end{aligned}
$$

where

$$
(2.3) \qquad a(X_t) = \frac{1}{2} c(X_t) c(X_t)^T
$$

is a semi-positive definite matrix.

**Lemma 2.1.** *If we define $N'(\cdot, \cdot)$ as a random measure on $(\mathbb{R}_+, J)$ by*

$$
(2.4)\quad N'([0,t], B, \omega) = \int_0^t \int_E N(ds, dz, \omega) 1_{\{z:\ k(X_{s-}(\omega), z) \in B\}}, \quad \forall B \in \mathcal{J},\quad t > 0,
$$

*then $N'(dt, d\xi)$ has a compensator $L(X_{t-}, d\xi) dt$, where $L(\cdot, \cdot)$ is a transition kernel from $(D, \mathcal{D})$ to $(J, \mathcal{J})$, which is defined by*

$$
(2.5) \qquad L(x, B) = \psi\{z \in E : k(x, z) \in B\}, \quad \forall B \in \mathcal{J}.
$$

*Furthermore, $L(\cdot, \cdot)$ can be rewritten as*

$$
(2.6) \qquad L(x, d\xi) = \lambda(x) Q(x, d\xi),
$$

*where $\lambda(x)$ represents the jump intensity of the process $X$, and $Q(x, \cdot)$ is the jump measure that is a probability kernel for the distribution of jump magnitudes.*

*Proof.* For each $B \in \mathcal{J}$ and $s, t \in \mathbb{R}_+$, we have

$$
\begin{aligned}
\mathbb{E}_x & \left[ \int_0^{s+t} (N'(du, B) - du L(X_{u-}, B)) \Big| \mathcal{F}_s \right] \\
& = \int_0^s (N'(du, B) - du L(X_{u-}, B)) \\
& \quad + \mathbb{E}_{X_s} \left[ \int_s^{s+t} \left( \int_E N(du, dz) 1_{\{z:\ k(X_{u-}, z) \in B\}} - du L(X_{u-}, B) \right) \right] \\
& = \int_0^s [N'(du, B) - du L(X_{u-}, B)].
\end{aligned}
$$

By the monotone class theorem, we can deduce that for each $f \in \mathcal{J}_+$,

$$
M_t^f := \int_0^t \int_J [N'(du, d\xi) - du L(X_{s-}, d\xi)] f(\xi),
$$

is a martingale, and since $L(X_{t-}, \cdot)$ is predictable, the first part of Lemma 2.1 has been proved.

Since $\psi(E) < \infty$, then $L(x, J) < \infty$. Therefore by simply defining $\lambda(x) = L(x, J)$ we obtain

$$Q(x, \cdot) = \begin{cases} \frac{L(x, \cdot)}{\lambda(x)} & \lambda(x) > 0 \\ \delta_x(\cdot) & \lambda(x) = 0, \end{cases} \tag{2.7}$$

for each $x \in D$, where $\delta_x(\cdot)$ denotes the Dirac measure at $x$. This completes the proof. $\square$

Now using (2.2) and Lemma 1.1, we can derive the following consistency condition for a jump diffusion model.

**Theorem 2.1.** *Under Assumption 2.1, a multi-factor jump-diffusion model $\mathcal{M}$ is consistent if, and only if, for each $(\tau, x) \in \mathbb{R}_+ \times D$, given the forward rates curve $G(\tau, X_t)$, the coefficients $a(x)$, $b(x)$, $\lambda(x)$ and $Q(x, \cdot)$ satisfy the following constraint.*

$$-\partial_\tau G(\tau, x) + \sum_{i=1}^n \partial_{x_i} G(\tau, x) b_i(x) + \sum_{i,j=1}^n a_{i,j}(x) \partial_{x_i} \partial_{x_j} G(\tau, x) \tag{2.8}$$

$$= 2 \sum_{i,j=1}^n a_{i,j}(x) \partial_{x_i} G(\tau, x) \int_0^\tau \partial_{x_j} G(u, x) du$$

$$- \lambda(x) \int_J \delta_0(x, \tau, \xi) Q(x, d\xi),$$

*where $\delta_0(x, \tau, \xi) = [G(\tau, x + \xi) - G(\tau, x)] e^{-\int_0^\tau (G(u, x+\xi) - G(u, x)) du}$.*

*Proof.* Using Lemma 1.1 and Lemma 2.1, it follows from (2.2) that

$$\sum_{i=1}^n \int_0^t \partial_{x_i} G(\tau, X_{s-}) b_i(X_{s-}) ds + \sum_{i,j=1}^n \int_0^t a_{i,j}(X_{s-}) \partial_{x_i} \partial_{x_j} G(\tau, X_{s-}) ds \tag{2.9}$$

$$= 2 \sum_{i,j=1}^n \int_0^t a_{i,j}(X_{s-}) \partial_{x_i} G(\tau, X_{s-}) \int_0^\tau \partial_{x_j} G(u, X_{s-}) du \, ds$$

$$- \int_0^t ds \int_J \delta_0(X_{s-}, \tau, \xi) L(X_{s-}, d\xi) + \int_0^t \partial_\tau G(\tau, X_s) ds.$$

Notice that there are at most countably many jumps for the process $X$ during the time 0 to $t$, for each $t > 0$. Therefore, without loss of generality, we do not distinguish between $X_{s-}$ and $X_s$ in (2.9). By Assumption 2.1, we can obtain (2.8) by differentiating both sides of (2.9) with respect to $t$. $\square$

**Assumption 2.2.** *The jump intensity $\lambda(\cdot)$ is a continuous function on $D$, and the probability kernel $Q(x, \cdot)$ defined by (2.7) does not depend on $x$, i.e.,*

$$L(x, d\xi) = \lambda(x) Q(d\xi). \tag{2.10}$$

**Remark 2.1.** *The models with the jump measure defined by (2.10) include two specific models: pure diffusion models ($\lambda(\cdot) \equiv 0$), and models driven by Lévy processes or more precisely, compound Poisson processes when the intensity $\lambda$ is independent of the state $X$.*

Now we can derive the following characterization theorem for a multi-factor jump-diffusion model $\mathcal{M}$.

**Theorem 2.2.** *Under Assumptions 2.1 and 2.2, the jump measure $Q(\cdot)$ can be freely chosen subject only to the regularity condition:*

$$(2.11) \qquad \int_J \delta_0(x, \tau, \xi) Q(d\xi) < \infty, \quad \forall \, (\tau, x) \in \mathbb{R}_+ \times D.$$

*Furthermore, if the forward rate curve $G(\cdot, x)$ satisfies the condition that the functions $\partial_{x_i} G(\cdot, x)$, $\partial_{x_i} \partial_{x_j} G(\cdot, x)$, $\int_J \delta_0(x, \cdot, \xi) Q(d\xi)$ and $\partial_{x_i} G(\cdot, x) \int_0^\cdot \partial_{x_j} G(u, x) du$ are linearly independent for all $1 \leq i, j \leq n$ and for all $x$ in some dense subset $D_0 \subset D$, then the drift $a(\cdot)$, diffusion $b(\cdot)$ and jump intensity $\lambda(\cdot)$ of the state process $X$ are uniquely determined by $G$.*

*Proof.* Set $M = (n+1)^2$ and choose a sequence $0 \leq \tau_1 < \tau_2 < \cdots < \tau_M$, such that by the linear independence condition, the $M \times M$ matrix with the $i$th row constructed as

$$\Big(\partial_{x_1} G(\tau_i, x), ..., \partial_{x_n} G(\tau_i, x), \partial_{x_1}\partial_{x_1} G(\tau_k, x), ..., \partial_{x_n}\partial_{x_n} G(\tau_i, x),$$

$$\partial_{x_1} G(\tau_i, x) \int_0^{\tau_i} \partial_{x_1} G(u, x) du, ..., \partial_{x_n} G(\tau_i, x) \int_0^{\tau_i} \partial_{x_n} G(u, x) du, \int_J \delta_0(x, \tau_i, \xi) Q(d\xi)\Big),$$

for each $i = 1, 2, ..., M$, is invertible. Therefore $a(x)$, $b(x)$ and $\lambda(x)$ are uniquely determined by the arbitrage-free condition (2.8), for each $x \in D_0$. Because of the continuity of $a$, $b$ and $\lambda$, the extensions to the state space $D$ are unique. This completes the proof of Theorem 2.2. □

Now by applying Theorem 2.2, we can discuss several specific cases. For simplicity, throughout the following sections Assumptions 2.1 and 2.2 are satisfied.

## 3. Separable Term Structure Models

In this section, we consider the case in which the forward rate curve $G(\tau, x)$ has a separable structure[1]

$$(3.1) \qquad G(\tau, x) = \sum_{k=1}^m h_k(\tau) \phi_k(x),$$

where $h_k : \mathbb{R}_+ \mapsto \mathbb{R}$ is $C^1$, and $\phi_k \mapsto \mathbb{R}$ is $C^2$ for $k = 1, ..., m$. Therefore according to Theorem 2.1, we have the following consistency conditions.

---

[1]This class of models has been investigated by Filipović [15] in the diffusion case.

**Proposition 3.1.** *A separable term structure model (3.1) is consistent if, and only if, the following equation holds:*

$$(3.2) \sum_{k=1}^{m}(h_k(\tau) - h_k(0))\phi_k(x) = \sum_{i=1}^{n}\Gamma_i(\tau,x)b_i(x)$$
$$+ \sum_{i,j=1}^{n} a_{i,j}(x)(\Lambda_{i,j}(\tau,x) - 2\Gamma_i(\tau,x)\Gamma_j(\tau,x))$$
$$+ \lambda(x)\Psi(H(\tau),x), \quad \forall\, (\tau,x) \in \mathbb{R}_+ \times D,$$

*where for $\forall\, 1 \leq i,j \leq n$ and $v \in \mathbb{R}^m$,*

$$(3.3) \qquad \Gamma_i(\tau,x) := \sum_{k=1}^{m} H_k(\tau)\frac{\partial \phi_k(x)}{\partial x_i},$$

$$(3.4) \qquad \Lambda_{i,j}(\tau,x) := \sum_{k=1}^{m} H_k(\tau)\frac{\partial^2 \phi_k(x)}{\partial x_i \partial x_j},$$

$$(3.5) \qquad \Psi(v,x) := \int_J \left(1 - e^{-\langle v, \phi(x+\xi) - \phi(x)\rangle}\right) Q(d\xi)$$

$$(3.6) \qquad \text{with} \qquad H_k(\tau) := \int_0^\tau h_k(u)du,$$

$$(3.7) \qquad H(\tau) := (H_1(\tau), H_2(\tau), ..., H_m(\tau))^T,$$

$$(3.8) \qquad \phi(x) := (\phi_1(x), \phi_2(x), ..., \phi_m(x))^T.$$

*Moreover, if we assume that the functions*

$$\Psi(H(\cdot),x), \quad \Gamma_i(\cdot,x), \quad \Lambda_{i,j}(\cdot,x) - 2\Gamma_i(\cdot,x)\Gamma_j(\cdot,x), \quad \forall\, 1 \leq i,j \leq n$$

*are linearly independent for all $x$ in some dense subset $D_0 \subset D$, then $a$, $b$, and $\lambda$ are uniquely determined by $h$, $\phi$ and the measure $Q$.*

*Proof.* The result follows from Theorems 2.1 and 2.2. □

Now on setting

$$B_k(x) := \sum_{i=1}^{n} b_i(x)\frac{\partial \phi_k(x)}{\partial x_i} + \sum_{i,j=1}^{n} a_{i,j}(x)\frac{\partial^2 \phi_k(x)}{\partial x_i \partial x_j},$$

$$\text{and} \quad A_{k,l}(x) = A_{l,k}(x) := 2 \sum_{i,j=1}^{n} a_{i,j}(x)\frac{\partial \phi_k(x)}{\partial x_i}\frac{\partial \phi_l(x)}{\partial x_j}, \quad \forall 1 \leq k,l \leq m,$$

it follows from (3.2) that

$$(3.9) \quad \sum_{k=1}^{m}(h_k(\tau) - h_k(0))\phi_k(x) = \sum_{k=1}^{m} H_k(\tau)B_k(x) - \sum_{k,l=1}^{m} H_k(\tau)H_l(\tau)A_{k,l}(x)$$
$$+ \lambda(x)\Psi(H(\tau),x), \quad \forall\, (\tau,x) \in \mathbb{R}_+ \times D.$$

Therefore, once we know $(a(\cdot), b(\cdot), \lambda(\cdot), Q(\cdot), (h_i(0)_{0 \leq i \leq n})$ and $\phi(\cdot)$, we can derive the representation for $H(\cdot)$. This is clarified by the following proposition.

**Proposition 3.2.** *Suppose that the functions $\phi_1, ..., \phi_m$ are linearly independent. Then the coefficient functions $H_1, ..., H_m$ solve a system of ordinary differential equations (ODEs):*

$$\frac{dH_k(\tau)}{d\tau} = R_k(H(\tau)), \quad 1 \leq k \leq m, \tag{3.10}$$

*where $R_k$ has the form*

$$R_k(v) = \theta_k + \langle \beta_k, v \rangle - \langle \alpha_k v, v \rangle + \gamma_k(v), \quad \forall v \in \mathbb{R}^m \tag{3.11}$$

*with $\theta_k = h_k(0)$, $\beta_k \in \mathbb{R}^m$, $\alpha_k$ is an $m \times m$ symmetric matrix, and $\gamma_k(v)$ is a linear combination of terms of the form $\Psi(v, x)$, where $\Psi(v, x)$ is introduced in (3.5).*

*Proof.* Choose $m$ mutually distinct points $(x^l)_{1 \leq l \leq m}$ in $D$ such that the $m \times m$ matrix $(\phi_k(x^l))$ is invertible. Then by multiplying the inverse matrix on both sides of (3.9) and setting $\theta_k = h_k(0)$, we obtain (3.10). It is easy to see $R_i(\cdot)$ has the form of (3.11) by appropriately setting $\alpha_k$, $\beta_k$ and $\gamma_k$. □

**Remark 3.1.** *Notice that specifying $(a(\cdot), b(\cdot), \lambda(\cdot), Q(\cdot), (h_k(0)_{0 \leq k \leq m}), (\phi_k(\cdot)_{0 \leq i \leq m}))$ is equivalent to specifying a multi-factor short rate model. Therefore Proposition 3.2 provides a way to solve the forward rate structure by applying the consistency requirements. Moreover it implies a necessary condition for a model to be consistent: the existence of the solution of the ODE system (3.10).*

3.1. **Affine Term Structure Models.** Now we will take a look at the simplest class of separable term structure models, namely the affine term structure models. The next proposition gives necessary and sufficient condition for a jump-diffusion model to be affine.

**Proposition 3.3.** *If the functions $a(x), b(x), \lambda(x), G(0, x)$ are affine, and $Q(\cdot)$ satisfies (2.11), then the term structure of forward rates $G(\cdot, x)$ is affine, that is,*

$$G(\tau, x) = h_0(\tau) + \sum_{i=1}^{n} h_i(\tau) x_i, \quad \forall (\tau, x) \in \mathbb{R}_+ \times D, \tag{3.12}$$

*for a continuously differentiable function $h$. Conversely if $G(\tau, x)$ is affine and*

$$H_1, ..., H_n, H_1 H_1, H_1 H_2, ..., H_n H_n, 1 - \Psi(H)$$

*are linearly independent functions, then the functions $a(\cdot), b(\cdot)$ and $\lambda(\cdot)$ are affine.*

*Proof.* The first part of the proposition is well established (e.g. [5], [12]). In particular, one can show that given a jump-diffusion model with the drift, diffusion, jump intensity and the short rate being affine functions of the state, the price of a zero-coupon bond has an exponential affine form and the coefficient functions solve a series of generalized Riccati equations. Thus the term structure is affine.

For the second part of the proof we will use Proposition 3.1. If we set

$$\phi_0(0) = 1, \quad \phi_k(x) = x_k, \quad 1 \leq k \leq n,$$

in (3.2) then we can derive the following consistency condition for affine term structure models:

(3.13)
$$h_0(\tau) - h_0(0) + \sum_{i=1}^{n}(h_i(\tau) - h_i(0))x_i = \sum_{i=1}^{n} H_i(\tau)b_i(x) - \sum_{i,j=1}^{n} a_{i,j}(x)H_i(\tau)H_j(\tau) + \lambda(x)(1 - \Psi(H(\tau))),$$

where $H(\cdot) = (H_1(\cdot), ..., H_n(\cdot))^T$ and $\Psi(v) = \int_J e^{-\langle v,\xi \rangle} Q(d\xi)$, which is the Laplace transform of the probability measure $Q$. Since the left hand side of (3.13) is affine and the coefficient matrix is invertible and independent of $x$, $a(x)$, $b(x)$ and $\lambda(x)$ must be affine functions of $x$. This completes the proof. □

Now it is assumed that $a(\cdot)$, $b(\cdot)$, $\lambda(\cdot)$ and $Q(\cdot)$ are given as affine functions, and $(h_i(0))_{0 \leq i \leq n}$ are known. The following corollary can be derived directly from Proposition 3.2.

**Corollary 3.1.** *Under the consistency condition (3.13), if $a(\cdot)$, $b(\cdot)$, $\lambda(\cdot)$ and $Q(\cdot)$ satisfy Assumptions 2.1 and 2.2, the coefficient functions $(H_i(\cdot))_{0 \leq i \leq n}$ can be determined from a system of ODEs shown as follows. For $k = 0, 1, ...n$,*

(3.14)
$$\frac{dH_k(\tau)}{d\tau} = R_k(H(\tau)),$$

*where $R_k$ has the form*

(3.15) $R_k(v) = \theta_k + \langle \beta_k, v \rangle + \langle \alpha_k v, v \rangle + \gamma_k \int_J \left(1 - e^{-\sum_{j=1}^{n} v_j \xi_j}\right) Q(d\xi),$

*with $\theta_k = h_k(0)$.*

**Remark 3.2.** *The above system of ODEs (3.14) and (3.15) are known as generalized Riccati equations (GREs), which have also been derived in [10] using a different approach. The existence and uniqueness of the solutions to the GREs have also been studied in [10].*

Under regularity conditions, Filipović [15] demonstrated that affine and quadratic models are the only two possible consistent models that can produce separable polynomial term structure in the diffusion case. We have already seen that affine term structure models allow for jumps in the state process $X$. However, importing jumps into the state process is generally difficult with polynomial term structure models[2] if not impossible. For example, it was shown in [7] that with a quadratic term structure, the state process $X_t$ can only follow a so-called quadratic process that does not allow jumps. For this reason, several alternative approaches have been adopted by researchers. One approach is to accommodate the existing affine jump-diffusion and quadratic models into one framework, namely the linear-quadratic jump-diffusion (LQJD) class originally proposed by Piazzesi [20] and further developed by Cheng and Scaillet [8]. Another approach is to apply a special Lévy process to drive the dynamics of the state variables (see [2]). Then pricing bonds and other derivatives can be achieved by approximation (see [6], [17]).

---

[2]This means that $\phi_k(x)$ defined in (3.1) is a polynomial in $x$, for each $1 \leq k \leq m$.

## 4. The Nelson-Siegel Curves

In this section, we discuss a popular non-separable term structure model, namely the Nelson-Siegel curve family (see [18]). This curve family has been studied in [13], and as we will show, there exists no non-trivial diffusion model consistent with the Nelson-Siegel forward curve.

The Nelson-Siegel forward curves are given by the form:

$$(4.1) \qquad G(\tau, x) = x_1 + (x_2 + x_3\tau)e^{-x_4\tau}, \quad (x_1, x_2, x_3) \in \mathbb{R}^3, \quad x_4 > 0.$$

Since $X^4 > 0$, we restrict the support of the jump measure $Q(d\xi)$ to $\{\xi_4 \geq 0\}$. Thus we redefine $D := \mathbb{R}^3 \times \mathbb{R}_{++}$ and $J := \mathbb{R}^3 \times \mathbb{R}_+$ in this section. We also make the following regularity assumption.

**Assumption 4.1.** *The measure $Q$ of the jump magnitude satisfies the following regularity condition:*

$$(4.2) \qquad \int_J (1 + |\xi_4|^3) e^{-r_2\xi_2 - r_3\xi_3} Q(d\xi) < \infty, \quad \forall\, r_2, r_3 \in \mathbb{R}_+.$$

If we move the term $2\sum_{i,j=1}^n a_{i,j}(x)\partial_{x_i} G(\tau, x) \int_0^\tau \partial_{x_j} G(u, x) du$ to the LHS of (2.8), then using (A.19)-(A.23) of the Appendix, we can write the consistency condition for the Nelson-Siegel curve as

$$(4.3) \qquad q_0(\tau, x) + q_1(\tau, x)e^{-x_4\tau} + q_2(\tau, x)e^{-2x_4\tau} = \lambda(x) \int_J \delta_0(x, \tau, \xi) Q(d\xi),$$

where $q_0(\tau, \cdot)$, $q_1(\tau, \cdot)$ and $q_2(\tau, \cdot)$ are polynomial functions of $\tau$ with degrees less than or equal to 1, 3, and 4, respectively (shown in the Appendix).

The following theorem asserts that under regularity conditions, there exists no non-trivial multi-factor model consistent with the Nelson-Siegel forward curve in the jump-diffusion case. The proof is included in the Appendix.

**Theorem 4.1.** *Under Assumptions 2.1, 2.2 and 4.1, a jump-diffusion model with the Nelson-Siegel-type forward curves is consistent if, and only if, the state process $X$ is deterministic upon $x$; i.e., for $\forall\, x \in D$, $a(x) = 0$, and either $\lambda(x) = 0$ or the jump measure $Q$ is the Dirac measure at 0.*

## 5. Conclusion

Motivated by previous work on consistency problems for diffusion models, this article investigated this issue in the jump-diffusion case. Unlike the diffusion case, here we have four elements to consider: the drift, diffusion, jump intensity and jump size measure. This difference seems to give us more freedom for making a model consistent. We have shown that the jump size measure $Q$ can be chosen freely, and once the jump size measure is given, under some regularity conditions, the drift, diffusion and jump intensity are uniquely determined in terms of $G$ by the consistency requirement.

For separable term structure models, in addition to the consistency condition given by Proposition 3.1, we also derive a necessary condition which is given by the existence of a solution of the ODEs (3.10). This indicates that, given the short rate model one can solve for the term structure of the forward rates from these ODEs. Accordingly the price of a zero-coupon bond can be derived by using (1.2).

It is well known that there exists no non-trivial diffusion model consistent with the Nelson-Siegel-type forward curve. We have demonstrated that this result is also true under technical assumptions in the jump-diffusion case.

## A. Appendix

### A.1. The Proof of Theorem 4.1.

For any $x \in D$ such that $\lambda(x) > 0$, we can divide both sides of (4.3) by $\lambda(x)$ to obtain

$$(A.1) \qquad p_0(\tau, x) + p_1(\tau, x)e^{-x_4\tau} + p_2(\tau, x)e^{-2x_4\tau} = \int_J \delta_0(x, \tau, \xi) Q(d\xi),$$

where

$$(A.2) \qquad p_i(\tau, \cdot) = \frac{q_i(\tau, \cdot)}{\lambda(x)}, \quad i = 0, 1, 2.$$

Then the expectation of $\delta_0$ under the measure $Q$ has the form:

$$(A.3) \qquad \delta(\tau, x) = p_0(\tau, x) + p_1(\tau, x)e^{-x_4\tau} + p_2(\tau, x)e^{-2x_4\tau},$$

where

$$\begin{aligned} \delta(\tau, x) &= \mathbb{E}^Q[\delta_0(x, \tau, \xi)] \\ &= \int_J \delta_0(x, \tau, \xi) Q(d\xi). \end{aligned}$$

Since

$$(A.4) \qquad \delta_0(x, \tau, \xi) = [G(\tau, x+\xi) - G(\tau, x)]e^{-\int_0^\tau (G(u, x+\xi) - G(u, x))du},$$

it follows from Fubini's theorem that

$$(A.5) \qquad 1 - \int_0^\tau \delta(u, x) du = \int_J e^{-\int_0^\tau (G(u, x+\xi) - G(u, x))du} Q(d\xi).$$

By substituting (A.3) into (A.5), we see that the LHS of (A.5) should have the following form:

$$(A.6) \qquad p_0'(\tau, x) + p_1'(\tau, x)e^{-x_4\tau} + p_2'(\tau, x)e^{-2x_4\tau},$$

where $p'_0(\tau, \cdot), p'_1(\tau, \cdot)$ and $p'_2(\tau, \cdot)$ are polynomials in $\tau$ with degrees less than or equal to 2, 3, and 4, respectively. Now by (A.5) and (A.6), we can write

$$\lim_{\tau \to \infty} \frac{1}{\tau^3} \int_J e^{-\int_0^\tau (G(u,x+\xi) - G(u,x))du} Q(d\xi) = 0. \tag{A.7}$$

On the other hand, by Fatou's Lemma, we have

$$\lim_{\tau \to \infty} \frac{1}{\tau^3} \int_J e^{-\int_0^\tau (G(u,x+\xi) - G(u,x))du} Q(d\xi) \geq \tag{A.8}$$

$$\int_J \lim_{\tau \to \infty} \frac{1}{\tau^3} e^{-\int_0^\tau (G(u,x+\xi) - G(u,x))du} Q(d\xi).$$

By (A.24), we see that the RHS of (A.8) increases without bound unless the support of the measure $Q$ is restricted to the set $\{\xi_1 \geq 0\}$. Therefore in order to satisfy (A.3), we have proved that the jump size measure $Q$ has support restricted to $\{\xi_1 \geq 0\}$. Then it is straightforward to deduce that, for each $(\tau, x, \xi) \in \mathbb{R}_+ \times D \times J \cap \{\xi_1 \geq 0\}$, we have

$$e^{-\int_0^\tau (G(u,x+\xi) - G(u,x))du} \leq C_0(x) \max\left\{ e^{-\frac{x_2+\xi_2}{x_4+\xi_4} - \frac{x_3+\xi_3}{(x_4+\xi_4)^2}}, e^{-\frac{x_3+\xi_3}{(x_4+\xi_4)^2}}, 1 \right\}. \tag{A.9}$$

Now since (4.2) guarantees that the RHS of (A.9) is integrable, by the dominated convergence theorem, we have that

$$\lim_{\tau \to \infty} \int_J e^{-\int_0^\tau (G(u,x+\xi) - G(u,x))du} Q(d\xi) \tag{A.10}$$

$$= \lim_{\tau \to \infty} \int_{J \cap \{\xi_1 \geq 0\}} e^{-\int_0^\tau (G(u,x+\xi) - G(u,x))du} Q(d\xi)$$

$$= \int_{J \cap \{\xi_1 \geq 0\}} \lim_{\tau \to \infty} e^{-\int_0^\tau (G(u,x+\xi) - G(u,x))du} Q(d\xi)$$

$$\leq \int_{J \cap \{\xi_1 \geq 0\}} e^{-\frac{x_2+\xi_2}{x_4+\xi_4} - \frac{x_3+\xi_3}{(x_4+\xi_4)^2} + \frac{x_3}{x_4^2}} Q(d\xi)$$

$$< \infty.$$

This implies that $p'_0(\tau, x) = p'_0(x)$, which is independent of $\tau$. Therefore we can conclude that in (A.3), $p_0(\tau, x) = 0$, which indicates

$$\lim_{\tau \to \infty} \frac{e^{x_4 \tau} \delta(\tau, x)}{\tau^4} = 0. \tag{A.11}$$

By (A.3), we have

$$\frac{e^{x_4 \tau} \delta(\tau, x)}{\tau^4} = \int_{\{\xi_1 \geq 0\}} \frac{1}{\tau^4} \left( \xi_1 e^{(x_4 - \xi_1)\tau} + g(x, \xi, \tau) \right) f(x, \xi, \tau) Q(d\xi), \tag{A.12}$$

where

$$f(x, \xi, \tau) = e^{-\int_0^\tau (G(u,x+\xi) - G(u,x))du + \xi_1 \tau} > 0,$$

$$\text{and} \quad g(x, \xi, \tau) = [(x_2 + \xi_2) + (x_3 + \xi_3)\tau] e^{-(\xi_1 + \xi_4)\tau} - (x_2 + x_3 \tau) e^{-\xi_1 \tau}.$$

By (4.2) and (A.9), we have

$$\text{(A.13)} \quad \lim_{\tau \to \infty} \int_{J \cap \{\xi_1 \geq 0\}} \frac{1}{\tau^4} g(x, \xi, \tau) f(x, \xi, \tau) Q(d\xi) = 0.$$

Thus, we find that

$$\text{(A.14)} \quad \lim_{\tau \to \infty} \int_{J \cap \{\xi_1 \geq 0\}} \frac{1}{\tau^4} \xi_1 e^{(x_4 - \xi_1)\tau} f(x, \xi, \tau) Q(d\xi) = 0, \quad \forall\, x \in D.$$

Since

$$\lim_{\tau \to \infty} f(x, \xi, \tau) = e^{-\frac{x_2 + \xi_2}{x_4 + \xi_4} - \frac{x_3 + \xi_3}{(x_4 + \xi_4)^2} + \frac{x_2}{x_4} + \frac{x_3}{x_4^2}} > 0,$$

it follows from Fatou's Lemma that

$$\begin{aligned} 0 &= \lim_{\tau \to \infty} \int_{J \cap \{\xi_1 \geq 0\}} \frac{1}{\tau^4} \xi_1 e^{(x_4 - \xi_1)\tau} f(x, \xi, \tau) Q(d\xi) \\ &\geq \int_{J \cap \{\xi_1 \geq 0\}} \lim_{\tau \to \infty} \frac{1}{\tau^4} \xi_1 e^{(x_4 - \xi_1)\tau} f(x, \xi, \tau) Q(d\xi) \\ &\geq \int_{J \cap \{0 < \xi_1 < x_4\}} \lim_{\tau \to \infty} \frac{1}{\tau^4} \xi_1 e^{(x_4 - \xi_1)\tau} f(x, \xi, \tau) Q(d\xi). \end{aligned}$$

Thus (A.14) holds only if the support of the measure $Q$ is restricted to $\{\xi_1 = 0\} \cup \{\xi_1 \geq x_4\}$. Therefore the consistency condition becomes

$$\text{(A.15)} \quad p_1(\tau, x) + p_2(\tau, x) e^{-x_4 \tau} = \int_{J \cap \{\xi_1 = 0\}} A_1(\tau, x, \xi) Q(d\xi) + \int_{J \cap \{\xi_1 \geq x_4\}} A_2(\tau, x, \xi) Q(d\xi),$$

where

$$\begin{aligned} A_1(\tau, x, \xi) &= g(x, \xi, \tau) f(x, \xi, \tau) \\ A_2(\tau, x, \xi) &= \left( \xi_1 e^{(x_4 - \xi_1)\tau} + g(x, \xi, \tau) \right) f(x, \xi, \tau). \end{aligned}$$

It is easy to see that

$$\lim_{\tau \to \infty} \frac{1}{\tau} \left[ \int_{J \cap \{\xi_1 = 0\}} A_1(\tau, x, \xi) Q(d\xi) + \int_{J \cap \{\xi_1 \geq x_4\}} A_2(\tau, x, \xi) Q(d\xi) \right] < \infty.$$

However

$$\lim_{\tau \to \infty} \frac{p_1(\tau, x)}{\tau} = \infty,$$

unless $p_1(\tau, x)$ is affine in $\tau$. Therefore $p_1(\tau, \cdot)$ is an affine function of $\tau$. Now we differentiate both sides of (A.15) twice. The condition (4.2) enables us to exchange the order of integration and differentiation, and thus we obtain

$$\text{(A.16)} \quad \tilde{p}_2(\tau, x) e^{-x_4 \tau} = \int_{J \cap \{\xi_1 = 0\}} B_1(\tau, x) Q(d\xi) + \int_{J \cap \{\xi_1 \geq x_4\}} B_2(\tau, x) Q(d\xi),$$

where $\tilde{p}_2(\tau, \cdot)$ is a polynomial function of $\tau$ with the degree less than 4,

$$B_1(\tau, x) = \frac{\partial^2 A_1(\tau, x)}{\partial \tau^2},$$

$$\text{and} \quad B_2(\tau, x) = \frac{\partial^2 A_2(\tau, x)}{\partial \tau^2}.$$

By multiplying $\frac{e^{x_4\tau}}{\tau^5}$ on both sides of (A.16), the LHS converges to 0 as $\tau \to \infty$; this holds for the RHS only if the support of the measure $Q$ is restricted to $\{\xi_1 \geq 2x_4\} \cup \{\xi_1 = 0, \xi_4 \geq x_4\} \cup \{\xi_1 = 0, \xi_4 = 0\}$.

Therefore (A.16) becomes

(A.17)
$$\tilde{p}_2(\tau, x)e^{-x_4\tau} = \int_{\{\xi_1=0,\xi_4=0\}} C_1(\tau, x, \xi)Q(d\xi) + \int_{\{\xi_1=0,\xi_4 \geq x_4\}} C_2(\tau, x, \xi)Q(d\xi)$$
$$+ \int_{J \cap \{\xi_1 \geq 2x_4\}} C_3(\tau, x, \xi)Q(d\xi),$$

where $C_1(\tau, x, \xi)$, $C_2(\tau, x, \xi)$ and $C_3(\tau, x, \xi)$ are given by (A.25)-(A.27), as derived below. On denoting the RHS of (A.17) by $R(\tau, x)$, it is easy to see that

$$\lim_{\tau \to \infty} \frac{R(\tau, x)e^{x_4\tau}}{\tau^2} = 0.$$

Thus we can conclude that $\tilde{p}_2(\cdot, x)$ is an affine function of $\tau$, and so are $p_2(\cdot, x)$ and $q_2(\cdot, x)$. By (A.2), we have already shown that $q_0(\cdot, x) = 0$, and the polynomials $q_1(\cdot, x)$ and $q_2(\cdot, x)$ are both affine in the first variable. Therefore by (A.28)-(A.41), as derived below, we conclude that

$$a_{11} = a_{33} = a_{44} = 0.$$

Since the diffusion matrix $a$ is semi-positive definite, this implies that all entries of $a$ are zero except $a_{22}$. Then we have that $q_2(\tau, x)$ and thus $p_2(\tau, x)$, and $\tilde{p}_2(\tau, x)$ are independent of $\tau$, and can thus be written as $q_2(x)$, $p_2(x)$, and $\tilde{p}_2(x)$ respectively. Thus by multiplying both sides of (A.17) by $e^{x_4\tau}$, it follows that the LHS is independent of $\tau$, but the RHS is a function of $\tau$ except for the case that $Q$ is the Dirac measure at 0. Therefore the consistency condition (4.3) can be rewritten as

(A.18) $\qquad q_0(\tau, x) + q_1(\tau, x)e^{-x_4\tau} + q_2(\tau, x)e^{-2x_4\tau} = 0, \quad \forall \tau > 0.$

As shown by Filipović [13], (A.18) implies that the diffusion matrix $a(x) = 0$.

Now when $\lambda(x) = 0$, we already have (A.18); thus we have shown that in order to satisfy the consistency condition (4.3), the diffusion matrix $a(\cdot) \equiv 0$. Moreover either $\lambda(x) = 0$ or the measure $Q$ is the Dirac measure at 0. Therefore we conclude that there exists no non-trivial jump-diffusion model consistent with the Nelson-Siegel family. This completes the proof of Theorem 4.1. □

A.2. **Some Derivations for the Nelson-Siegel Family.** By (4.1), it is straightforward to deduce that

$$G(\tau, x + \xi) - G(\tau, x) = \xi_1 + [(x_2 + \xi_2) + (x_3 + \xi_3)\tau] e^{-(x_4+\xi_4)\tau} \tag{A.19}$$
$$- (x_2 + x_3\tau)e^{-x_4\tau},$$

$$\partial_\tau G(\tau, x) = (x_3 - x_4 x_2 - x_3 x_4 \tau)e^{-x_4\tau}, \tag{A.20}$$

$$\nabla_x G(\tau, x) = (1, \quad e^{-x_4\tau}, \quad \tau e^{-x_4\tau}, \quad -\tau(x_2 + x_3\tau)e^{-x_4\tau})^T, \tag{A.21}$$

$$\nabla_x (\partial_{x_4} G(\tau, x)) = e^{-x_4\tau}(0, \quad -x_4, \quad -x_4\tau, \quad \tau^2(x_2 + x_3\tau))^T, \tag{A.22}$$

$$\text{and} \quad \frac{\partial^2}{\partial x_i \partial x_j} G(\tau, x) = 0, \quad 1 \leq i, j \leq 3. \tag{A.23}$$

We have the following expression for $\log f(x, \xi, \tau)$.

$$-\int_0^\tau (G(u, x+\xi) - G(u, x))du = -\xi_1 \tau + \frac{x_2 + \xi_2}{x_4 + \xi_4}(e^{-(x_4+\xi_4)\tau} - 1) \tag{A.24}$$
$$+ \frac{x_3 + \xi_3}{x_4 + \xi_4}e^{-(x_4+\xi_4)\tau}\tau + \frac{x_3 + \xi_3}{(x_4 + \xi_4)^2}e^{-(x_4+\xi_4)\tau}$$
$$- \frac{x_3 + \xi_3}{(x_4 + \xi_4)^2} + \frac{x_2}{x_4}(1 - e^{-x_4\tau}) - \frac{x_3}{x_4}e^{-x_4\tau}\tau$$
$$- \frac{x_3}{x_4^2}e^{-x_4\tau} + \frac{x_3}{x_4^2}.$$

Also, for $B_1(\tau, x, \xi)$ and $B_2(\tau, x, \xi)$, we have

$$B_1(\tau, x, \xi) = \{-2(x_3 + \xi_3)\xi_4 e^{-\xi_4\tau} + [(x_2 + \xi_2) + (x_3 + \xi_3)\tau]\xi_4^2 e^{-\xi_4\tau}$$
$$- 2\left[(x_3 + \xi_3)e^{-\xi_4\tau} + [(x_2 + \xi_2) - (x_3 + \xi_3)\tau]\xi_4 e^{-\xi_4\tau} - x_3\right]$$
$$\left[(x_3 + \xi_3)e^{-\xi_4\tau} + [(x_2 + \xi_2) + (x_3 + \xi_3)\tau]e^{-(x_4+\xi_4)\tau} - (x_2 + x_3\tau)e^{-x_4\tau}\right]$$
$$+ \left[[(x_2 + \xi_2) + (x_3 + \xi_3)\tau]e^{-(x_4+\xi_4)\tau} - (x_2 + x_3\tau)e^{-x_4\tau}\right]^3 e^{x_4\tau}\}$$
$$f(x, \xi, \tau).$$

and

$$B_2(\tau, x, \xi) = \{\xi_1 x_4^2 e^{x_4\tau} - 2(x_3 + \xi_3)\xi_4 e^{-\xi_4\tau} + [(x_2 + \xi_2) + (x_3 + \xi_3)\tau]\xi_4^2 e^{-\xi_4\tau}$$
$$- 2\left[\xi_1 x_4 e^{x_4\tau} + (x_3 + \xi_3)e^{-\xi_4\tau} + [(x_2 + \xi_2) - (x_3 + \xi_3)\tau]\xi_4 e^{-\xi_4\tau} - x_3\right]$$
$$\left[\xi_1 + (x_3 + \xi_3)e^{-\xi_4\tau} + [(x_2 + \xi_2) + (x_3 + \xi_3)\tau]e^{-(x_4+\xi_4)\tau} - (x_2 + x_3\tau)e^{-x_4\tau}\right]$$
$$+ \left[\xi_1 + [(x_2 + \xi_2) + (x_3 + \xi_3)\tau]e^{-(x_4+\xi_4)\tau} - (x_2 + x_3\tau)e^{-x_4\tau}\right]^3 e^{x_4\tau}\}$$
$$e^{-\xi_1\tau} f(x, \xi, \tau).$$

Similarly, $C_1(\tau, x, \xi)$, $C_2(\tau, x, \xi)$ and $C_3(\tau, x, \xi)$ are given by

$$C_1(\tau, x, \xi) = -2\xi_3[\xi_2 + \xi_3\tau]e^{-x_4\tau} + (\xi_2 + \xi_3\tau)^3 e^{-2x_4\tau} f(x, \xi, \tau). \tag{A.25}$$

(A.26)
$$C_2(\tau, x, \xi) = \{-2(x_3 + \xi_3)\xi_4 e^{-\xi_4 \tau} + [(x_2 + \xi_2) + (x_3 + \xi_3)\tau]\xi_4^2 e^{-\xi_4 \tau}$$
$$-2\left[(x_3 + \xi_3)e^{-\xi_4 \tau} + [(x_2 + \xi_2) + (x_3 + \xi_3)\tau]\xi_4 e^{-\xi_4 \tau} - x_3\right]$$
$$\left[(x_3 + \xi_3)e^{-\xi_4 \tau} + [(x_2 + \xi_2) - (x_3 + \xi_3)\tau]e^{-(x_4+\xi_4)\tau} - (x_2 + x_3\tau)e^{-x_4\tau}\right]$$
$$+ \left[[(x_2 + \xi_2) + (x_3 + \xi_3)\tau]e^{-(x_4+\xi_4)\tau} - (x_2 + x_3\tau)e^{-x_4\tau}\right]^3 e^{x_4\tau}\}$$
$$f(x, \xi, \tau).$$

and

(A.27)
$$C_3(\tau, x, \xi) = \{\xi_1 x_4^2 e^{x_4\tau} - 2(x_3 + \xi_3)\xi_4 e^{-\xi_4 \tau} + [(x_2 + \xi_2) + (x_3 + \xi_3)\tau]\xi_4^2 e^{-\xi_4 \tau}$$
$$-2\left[\xi_1 x_4 e^{x_4\tau} + (x_3 + \xi_3)e^{-\xi_4 \tau} + [(x_2 + \xi_2) + (x_3 + \xi_3)\tau]\xi_4 e^{-\xi_4 \tau} - x_3\right]$$
$$\left[\xi_1 + (x_3 + \xi_3)e^{-\xi_4 \tau} + [(x_2 + \xi_2) - (x_3 + \xi_3)\tau]e^{-(x_4+\xi_4)\tau} - (x_2 + x_3\tau)e^{-x_4\tau}\right]$$
$$+ \left[\xi_1 + [(x_2 + \xi_2) + (x_3 + \xi_3)\tau]e^{-(x_4+\xi_4)\tau} - (x_2 + x_3\tau)e^{-x_4\tau}\right]^3 e^{x_4\tau}\}$$
$$e^{-\xi_1 \tau} f(x, \xi, \tau).$$

Finally, $q_0(\tau, x)$, $q_1(\tau, x)$ and $q_2(\tau, x)$ in (4.3) are given by

(A.28) $$q_0(\tau, x) = q_0^0(x) + q_1^0(x)\tau$$

(A.29) $$q_1(\tau, x) = \sum_{i=0}^{3} q_i^1(x)\tau^i$$

(A.30) $$q_2(\tau, x) = \sum_{i=0}^{4} q_i^2(x)\tau^i,$$

where

$$\text{(A.31)} \quad q_0^0(x) = -b_1(x) + \frac{2a_{12}(x)}{x_4} + \frac{2a_{13}(x)}{x_4^2} - \frac{2x_2}{x_4^2} - \frac{4a_{14}(x)x_3}{x_4^3},$$

$$\text{(A.32)} \quad q_1^0(x) = 2a_{11}(x),$$

$$\text{(A.33)} \quad q_0^1(x) = -b_2(x) - x_2 x_4 + x_3 + \frac{2a_{22}(x)}{x_4} - \frac{2a_{12}(x)}{x_4} - \frac{2a_{24}(x)x_2}{x_4^2}$$
$$- \frac{2a_{24}(x)x_3}{x_4^3} - \frac{2a_{13}(x)}{x_4^2} + \frac{2a_{14}(x)x_2}{x_4^2} + \frac{4a_{14}(x)x_3}{x_4^3} + \frac{2a_{23}(x)}{x_4^2},$$

$$\text{(A.34)} \quad q_1^1(x) = -b_3(x) + b_4(x)x_2 - x_3 x_4 + 2a_{12}(x) + 2a_{24}(x) - \frac{2a_{13}(x)}{x_4}$$
$$+ \frac{2a_{33}(x)}{x_4^2} + \frac{4a_{34}(x)x_2}{x_4^2} + \frac{2a_{34}(x)x_3}{x_4^2} + \frac{2a_{44}(x)x_2^2}{x_4^2} + \frac{4a_{44}(x)x_2 x_3}{x_4^3}$$
$$+ \frac{2a_{14}(x)x_2}{x_4} + \frac{4a_{14}(x)x_3}{x_4^2} + \frac{2a_{23}(x)}{x_4^2} - \frac{2a_{24}(x)x_2}{x_4},$$

$$\text{(A.35)} \quad q_2^1(x) = b_4(x)x_3 - a_{44}(x)x_2 + 2a_{13}(x)$$
$$+ 2a_{34}(x) - 2a_{14}(x)x_2 - \frac{2a_{24}(x)x_3}{x_4} - \frac{2a_{34}(x)x_3}{x_4^2}$$
$$+ \frac{2a_{44}(x)x_2 x_3}{x_4^2} + \frac{4a_{44}(x)x_3^2}{x_4^3} + \frac{2a_{14}(x)x_3}{x_4},$$

$$\text{(A.36)} \quad q_3^1(x) = -a_{44}(x)x_3 - 2a_{14}(x)x_3,$$

$$\text{(A.37)} \quad q_0^2(x) = -\frac{2a_{22}(x)}{x_4} - \frac{2a_{23}(x)}{x_4^2} + \frac{2a_{24}(x)x_2}{x_4} + \frac{4a_{24}(x)x_3}{x_4^3},$$

$$\text{(A.38)} \quad q_1^2(x) = -\frac{4a_{23}(x)}{x_4} + \frac{4a_{24}(x)x_2}{x_4} + \frac{4a_{24}(x)x_3}{x_4^2} - \frac{2a_{33}(x)}{x_4^2}$$
$$- \frac{4a_{34}(x)x_2}{x_4^2} - \frac{4a_{34}(x)x_3}{x_4^3} - \frac{2a_{44}(x)x_2^2}{x_4^3} - \frac{4a_{44}(x)x_2 x_3}{x_4^3},$$

$$\text{(A.39)} \quad q_2^2(x) = \frac{4a_{24}(x)x_3}{x_4} - \frac{2a_{33}(x)}{x_4} - \frac{4a_{34}(x)x_2}{x_4} - \frac{6a_{34}(x)x_3}{x_4^2}$$
$$- \frac{2a_{44}(x)x_2^2}{x_4} - \frac{4a_{44}(x)x_2 x_3}{x_4^2} - \frac{2a_{44}(x)x_2 x_3}{x_4} - \frac{4a_{44}(x)x_3^2}{x_4^3},$$

$$\text{(A.40)} \quad q_3^2(x) = -\frac{4a_{34}(x)x_3}{x_4} - \frac{4a_{44}(x)x_2 x_3}{x_4} - \frac{4a_{44}(x)x_3^2}{x_4^2},$$

$$\text{(A.41) and} \quad q_4^2(x) = -\frac{2a_{44}(x)x_3^2}{x_4}.$$

Table 1. Summary of Notation

| Notation | Implications |
|---|---|
| $X$ | A jump-diffusion where the jumps follow a multivariate point process |
| $(D, \mathcal{D})$ | The state space $D \subset \mathbb{R}^n$ and its Borel $\sigma$-algebra $\mathcal{D} := \mathcal{B}(D)$ |
| $(J, \mathcal{J})$ | The jump space $J \subset \mathbb{R}^n$ and its Borel $\sigma$-algebra $\mathcal{J} := \mathcal{B}(J)$ |
| $C(D)$ | The Banach space of continuous functions on $D$ |
| $bD$ | The Banach space of bounded Borel-measurable functions on $D$ |
| $C_b(D)$ | The Banach space consisting of all bounded continuous functions on $D$ |
| $C^k(D)$ | The space of $k$-times differentiable functions $f$ on the interior of $D$ such that all partial derivatives of $f$ up to order $k$ are continuous |
| $\mathbb{R}_+$, $(\mathbb{R}_{++})$ | The set of positive (strictly positive) real numbers |
| $\mathcal{G}$ | The infinitesimal generator of $X$ |
| $Leb$ | The Lebesgue measure on $\mathbb{R}_+$ |
| $\langle \cdot, \cdot \rangle$ | The inner product in the vector space $\mathbb{R}^n$ |
| $\nabla f$ | The gradient of the function $f$ on $D$ |

Department of Mathematics, University of Michigan, 2074 East Hall, Ann Arbor, MI 48109-1109
*E-mail address*: `erhan@umich.edu`

Lehman Brothers, Fixed Income Derivatives Research, 745 Seventh Avenue, New York, NY 10019
*E-mail address*: `lichen@lehman.com`

Department of Electrical Engineering, Princeton University, Princeton, NJ 08544
*E-mail address*: `poor@princeton.edu`